\documentclass[12pt,preprint]{aastex}

\slugcomment{}

\shorttitle{ABUNDANCES OF HD~122563}
\shortauthors{HONDA et al.}

\begin{document}

\title{Neutron-capture elements in the very metal-poor star HD~122563\altaffilmark{1}}

\author{S. Honda\altaffilmark{2}, W. Aoki\altaffilmark{2}, Y. Ishimaru\altaffilmark{3}, S. Wanajo\altaffilmark{4}, S. G. Ryan\altaffilmark{5}}

\altaffiltext{2}{National Astronomical Observatory, Mitaka, Tokyo,
181-8588, Japan; e-mail: honda@optik.mtk.nao.ac.jp, aoki.wako@nao.ac.jp}
\altaffiltext{3}{Academic Support Center, Kogakuin University, Hachioji, Tokyo 192-0015, Japan; kt13121@ns.kogakuin.ac.jp}
\altaffiltext{4}{Research Center for the Early Universe, Graduate School of Science, University of Tokyo, Bunkyo-ku, Tokyo 113-8654, Japan; wanajo@resceu.s.u-tokyo.ac.jp}
\altaffiltext{5}{Department of Physics and Astronomy, The Open
University, Walton Hall, Milton Keynes, MK7 6AA, UK; present address: Centre for Astrophysics Research, STRI, University of Hertfordshire, College Lane, Hatfield AL10 9AB, United Kingdom; s.g.ryan@herts.ac.uk}
\altaffiltext{1}{Based on data collected at the Subaru Telescope,
which is operated by the National Astronomical Observatory of Japan.}

\begin{abstract}

We obtained high resolution, high S/N spectroscopy for the very
metal-poor star HD~122563 with the Subaru Telescope High Dispersion
Spectrograph.  
Previous studies have shown that this object has excesses of 
light neutron-capture elements, while its abundances of heavy ones
are very low.  In our spectrum covering 3070 -- 4780 \AA~ of this
object, 19 neutron-capture elements have been detected, 
including seven for the first time in this star 
(Nb, Mo, Ru, Pd, Ag, Pr, and Sm).
Upper limits are given for five other elements including Th. 
The abundance pattern shows a gradually decreasing trend, as a function 
of atomic number, from Sr to Yb, which is quite different from those 
in stars with excesses of r-process elements. 
This abundance pattern of neutron-capture elements provides new strong 
constraints on the models of nucleosynthesis responsible for the very 
metal-poor stars with excesses of light neutron-capture elements but without 
enhancement of heavy ones.

\end{abstract}


\section{Introduction}

The chemical composition of extremely metal-poor stars is expected to
reflect the yields from a quite small number of nucleosynthesis
processes.  Recent abundance analyses for extremely metal-poor stars
have provided quite valuable information on the origin of the
elements, 
\citep[in particular when combined with Galactic chemical evolution studies,
e.g.,][]{ishimaru99,ishimaru04},
and the individual nucleosynthesis processes involved.

The rapid neutron capture process (r-process) is known to be
responsible for about half of the abundances of elements heavier than
the iron-group in solar system material.  Although observational data
have been rapidly increasing in the past decade, the astrophysical 
site of the r-process is still unclear. 
Previous nucleosynthesis studies suggest several possibilities,
e.g., neutrino-driven winds \citep{woosley94, wanajo01} or prompt
explosions \citep{sumiyoshi01, wanajo03} of core-collapse
(Type~II/Ibc) supernovae, neutron star mergers \citep{Freiburghaus99},
jets from gamma-ray burst accretion discs \citep{Surman05}. 
All the scenarios proposed above involve, however, severe problems 
that remain to be solved, and no consensus has yet been achieved.

Models of the r-process nucleosynthesis are usually examined by 
comparison with the abundance pattern of the r-process component 
in solar-system material. 
Recent measurements for abundances of
neutron-capture elements in very metal-poor stars have been providing
useful constraints on these models. \citet{sneden96,sneden03} have
studied the chemical abundances of the extremely metal-poor star
CS~22892--052, the first example of a small but growing class of metal-poor
stars that exhibit very large excesses of r-process elements relative
to iron ([r-process/Fe] $> +1.0$).  An important result of their work
is that the relative abundance pattern of the neutron-capture elements
from the 2nd to the 3rd peak (56 $\leq$ {\it Z} $\leq$ 76) in this
star is identical, within observational errors, to that of the
(inferred) solar system r-process component.  This phenomenon is
sometimes referred to as ``universality'' of the r-process, having a large
impact on the studies of the nature of the r-process, and its
astrophysical site.

However, the abundance patterns of light neutron-capture elements (38
$\leq$ Z $\leq$ 48) in r-process enhanced stars exhibit clear
deviations from that of the solar system r-process component \citep{sneden00}.
This suggests the existence of another r-process site
which has contributed to the light r-process elements in solar-system
material.  
This process is sometimes called as ``weak r-process'', while the
process that is responsible for heavy neutron-capture elements (Ba
$\sim$ U) is referred to ``main r-process'' \citep[e.g., ]
[]{truran02, Wanajo05}.

Following the previous studies \citep[e.g., ][]{burris00,truran02}, \citet{honda04} 
and \citet{aoki05} have investigated the correlation of Sr and Ba abundances in very
metal-poor stars ([Fe/H] $< -2.5$) in some detail. 
They showed that the dispersion of the Sr abundances clearly decreases with
increasing Ba abundance. This correlation indicates the existence of
(at least) two processes: one enriches both heavy and light
neutron-capture elements, while the other yields light neutron-capture
elements with only small production of heavy ones. 
\citet{travaglio04} discuss the contribution of a primary process 
which has produced light neutron-capture elements in our Galaxy.
 
The abundance patterns produced by the main r-process have been studied
previously based on high resolution spectroscopy of metal-poor 
stars \citep[e.g., ][]{sneden96,sneden03,hill02} as mentioned 
above. By way of contrast, the abundance pattern of
neutron-capture elements produced by the weak r-process is still unclear. Since
the abundances of heavy neutron-capture elements yielded by this
process are very low, it is quite difficult to accurately measure their
abundances in metal-poor stars. In addition, most spectral lines of
light neutron-capture elements, in particular those with $41 \leq Z <
56$ exist in the near-UV region, in which atmospheric extinction makes
the observation with ground-based telescopes difficult.

We have been investigating light neutron-capture elements in very
metal-deficient stars from near-UV spectroscopy with the 8.2~m
Subaru Telescope. Ishimaru et al. (2006, in preparation) studied the
abundances of Pd and Ag, which exist between the first and second
abundance peaks produced by neutron-capture processes, for several
metal-poor stars. One of the targets is HD~122563, a well-studied
bright metal-poor ([Fe/H] $= -2.7$) giant. 
This object has very low abundances of heavy
neutron-capture elements \citep[e.g. {[Ba/Fe]} $= -1.21$][]{honda04}, 
while light neutron-capture elements show large excesses relative to heavy
ones (e.g. [Sr/Ba] $=+1.1$). 
Therefore, the abundance pattern of
neutron-capture elements in this star might well represent the yields of
the process producing light neutron-capture elements at low
metallicity. 

In this paper we report the abundance pattern of light and heavy
neutron-capture elements in HD~122563. 
In \S 2 we present our near-UV spectroscopy with Subaru/HDS, 
while we describe in detail the abundance analysis of neutron-capture 
elements in \S 3.
Our results are compared with the previous studies in \S 4.
We discuss the derived abundances of HD~122563, and the source 
of neutron-capture elements in \S5.
Finally, our conclusions are presented in \S 6.

\section{Observations and measurements}

High Dispersion Spectroscopy of HD~122563 was carried out with the 
Subaru Telescope High Dispersion Spectrograph \citep[HDS: ][]{nogu02} 
in 30 April, 2004. 
Our spectrum covers the wavelength range from 3070 to 4780 {\AA}
with a resolving power of $R$ = 90,000.  The total exposure time is 
5400 seconds (one 600~s exposure plus four 1200~s ones).  
The reduction was carried out in a standard manner
using the IRAF echelle package\footnote{IRAF is distributed by
National Optical Astronomy Observatories, which are operated by the Association
of Universities for Research in Astronomy, Inc., under cooperative agreement
with the National Science Foundation}. The signal to
noise ratio (S/N) of the spectrum (per 0.9 km s$^{-1}$ pixel) estimated 
from the photon counts is 140 at 3100 {\AA}; 480 at 3500 {\AA}; 860 
at 3900 {\AA}; 1080 at 4200 {\AA} and 1340 at 4500 {\AA}.

In order to measure Ba lines at 5853~{\AA} and 6141~{\AA}, a spectrum
obtained during the HDS commissioning covering 5090{\AA} -- 6410{\AA} was
used. The S/N ratio of this spectrum is 350 at 6000~{\AA}.

We adopted the line list used in previous studies for
line identifications and analyses of neutron-capture 
elements \citep[e.g.,][]{hill02,cowan02,sneden03,johnson04}.
In addition, the recent measurements of transition
probabilities and hyperfine splitting reported for some 
elements \citep[e.g., ][]{lawler01a,lawler01b,denhartog03} are incorporated.  
The line list used in the analysis is given in Table \ref{tab:table1-4}. \\
Equivalent widths of clean, isolated lines are measured by fitting 
gaussian profile.
The results are given in Table \ref{tab:table1-4}.

\section{Abundance analysis}

For our quantitative abundance measurements, we used the analysis
program SPTOOL developed by Y. Takeda, based on Kurucz's ATLAS9/WIDTH9
\citep{kurucz93}.  SPTOOL calculates synthetic spectra and equivalent
widths of lines on the basis of the given model atmosphere, line
data, and chemical composition, under the assumption of LTE.

We adopted the model atmosphere parameters (effective temperature: $T_{\rm eff}$, 
gravity: $g$, micro-turbulent velocity:  $v_{\rm turb}$, and metallicity: [Fe/H]) derived by \citet{honda04}: 
$T_{\rm eff}$ = 4570 K, $\log g$ = 1.1,  $v_{\rm turb}$ = 2.2 km s$^{-1}$, 
and [Fe/H] $= -2.77$. 
The estimated uncertainties are $\Delta T_{\rm eff}$ = 100 K, $\Delta \log g$ = 0.3 dex,
 $\Delta v_{\rm turb}$ = 0.5 km s$^{-1}$, and $\Delta$[Fe/H] = 0.19 dex (Honda et al. 2004).

The abundance analyses
were attempted for 24 neutron-capture elements from Sr to Th. The
effects of hyperfine and isotopic splitting are taken into account in
the analysis of Ba, La, Eu, and Yb (see references in Table 1). 

The derived abundances are given in Table \ref{tab:table2}. 
The abundances determined from individual lines are given in 
Table \ref{tab:table1-4}.
We used the solar system abundances obtained by \citet{asplund05} to 
derive [X/Fe] values.  
The size of the random errors are estimated
from the standard deviation (1$\sigma$) of the abundances derived
from individual lines for elements that have three or more lines
available for the abundance analysis.  
For the abundances of elements based only on one
or two lines, we employ the mean of the random errors estimated from
those elements with three or more lines available.

The uncertainties of model atmosphere
parameters result in systematic errors in the abundances of
neutron-capture elements.
The effects of these uncertainties on the abundance measurements 
are given in Table \ref{tab:error}.
The behavior of the errors is slightly different for abundances 
derived from neutral species and those from ionized ones.
This should be borne in mind in the discussion on the abundances of 
light neutron-capture elements, some of which were determined using lines of neutral species.
By contrast, such differences do not significantly affect our
discussion of the abundance ratios of heavy neutron-capture elements, 
which were all measured from ionized species.

The details of the analysis for individual neutron-capture elements 
are presented below.
For comparison purposes, the abundances of Cu and Zn were also determined.
In addition to the Cu 5105 {\AA} and Zn 4810 {\AA} lines, the UV 
lines (3250 - 3350 {\AA}) are also used in the abundance analyses 
(Table \ref{tab:table1-4}).

\subsection{Light neutron-capture elements ($38 \leq Z \leq 47$) }

Although the abundances of Sr, Y, and Zr have been determined by a number of
previous studies for HD~122563, measurements for other light
neutron-capture elements are quite limited so far.  We have newly
detected 5 elements (Nb, Mo, Ru, Pd, and Ag) in our spectrum, and derived
an upper limit for a sixth element (Rh).

Two strong lines of \ion{Sr}{2} (4077 and 4215{\AA}) and the weak line
of \ion{Sr}{1} (4607{\AA}) were measured.  The two \ion{Sr}{2}
resonance lines are so strong that the abundance derived from these
lines is sensitive to the microturbulent velocity and treatment of
damping. However, the agreement of the abundances derived from
\ion{Sr}{2} and \ion{Sr}{1} (see Table \ref{tab:table1-4}) suggests 
the reliability of our measurements.

\begin{figure}[p]
\includegraphics[width=10cm]{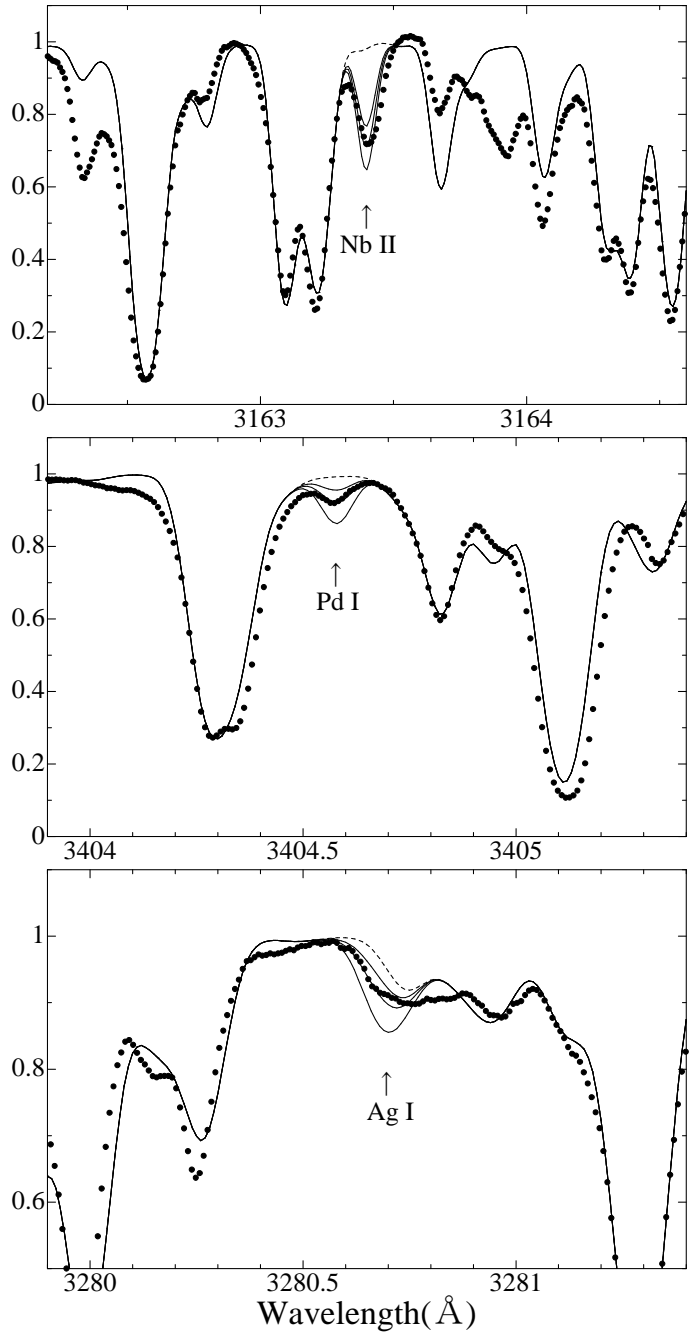}
\caption{The observed \ion{Nb}{2} 3163 {\AA}, \ion{Pd}{1} 3404 {\AA}, and 
\ion{Ag}{1} 3280 {\AA}. 
Dots: observations; 
solid lines: synthetic spectra computed for the adopted abundance (see Table 1) 
and values $\pm$0.3 dex different; 
dashed lines: synthetic spectra with no contribution from the line of interest.}

\label{fig:spectra}
\end{figure}

Two \ion{Nb}{2} ($Z=41$) lines are detected.  The detection of the 3163{\AA}
line is reported only for Canopus by \citet{reynolds88} and for the sun.  
This line would be useful for abundance
analyses because of its strength (Figure \ref{fig:spectra}). 
 We note, however, that the abundance determined from this line is 
slightly higher (by 0.25 dex) than that from
the 3215~{\AA} line, possibly suggesting a small contamination by other
spectral features.
However, here we simply adopt the mean of the abundances derived from 
the two lines.

The \ion{Mo}{1} ($Z=42$) 3864~{\AA} line is detected in CS~22892-052 
by \citet{sneden03}.
Although a blend of a CN line is reported
by these authors, that is not severe in HD~122563, because the
carbon abundance of this star is more than 1~dex lower than that of
CS~22892--052.

We have detected two lines of \ion{Ru}{1} ($Z=44$). 
The equivalent width of the 3498~{\AA} line is measurable due to there being
no (apparent) blend with other lines. 
Since a strong \ion{Fe}{1} line exists in the bluer
region of the other line at 3728~{\AA}, we applied the spectrum
synthesis technique to the analysis of that line.

The \ion{Rh}{1} ($Z=45$) 3692~{\AA} line exists in the wing of H I line, but is not
detected in our spectrum. We determined only an upper limit on the
abundance of this element.

We have detected two lines of Pd. The \ion{Pd}{1} 3242~{\AA} line
blends with OH lines.  The contamination was estimated using the OH
line list of \citet{kurucz93}, adjusting the oxygen abundance to match
the strengths of neighboring OH lines ([O/Fe] $= +$0.4). 
The \ion{Pd}{1} 3404~{\AA} line is clearly detected with no severe 
blend (Figure \ref{fig:spectra}).

Two \ion{Ag}{1} lines have been measured in some metal-poor stars. 
The \ion{Ag}{1} 3280~{\AA} line is detected, while the 3383~{\AA} line is not. 
Since NH lines blend with the 3280~{\AA} line \citep{johnson02}, 
we included the NH lines adopted from \citet{kurucz93} in the 
spectrum synthesis, calibrating the N abundance to reproduce the 
NH features around the \ion{Ag}{1} line ([N/Fe]$=-0.4$).
We note that the N abundance adopted is much lower than those derived by 
previous works \citep[e.g.,{[N/Fe]$=+1.1$}][]{westin00}.
This discrepancy is partially due to the inaccurate gf-values used 
in our analysis.
Recent studies of N abundances based on NH lines applied corrections 
of gf-values to the Kurucz's list \citep[e.g.,][]{ecuvillon04}.

In addition to the NH lines, some unidentified feature appears in the red part of 
this \ion{Ag}{1} line (Figure \ref{fig:spectra}). 
If such unknown lines also affect the \ion{Ag}{1} line, the values 
derived from this line must be regarded as an upper limit on the 
Ag abundance.

\subsection{Heavy neutron-capture elements ($56 \leq Z \leq 90$)}

The abundances of Ba, La, Eu, and Yb were derived by a spectrum
synthesis technique, taking into account hyperfine splitting. The
abundances of other elements were determined by applying a single line
approximation, which is justified by the fact that the absorption
lines are quite weak in general. 

In addition to the two resonance lines at 4554 and 4934~{\AA}, 
two Ba lines in the red region were measured.  
The line data of \citet{mcwilliam98} were used for the analysis 
of Ba lines, assuming the isotope ratios of r-process component 
in solar-system material. 
There is a small discrepancy between the Ba abundances from
the two resonance lines and others. 
A possible reason for this discrepancy is the non-LTE 
effect \citep[e.g.,][]{asplund05b}. 
However, we here simply adopt the
average of the Ba abundances from the four lines.

We detected three \ion{Eu}{2} lines at 3819 ~{\AA}, 4129~{\AA} and
4205~{\AA}.  However, the 4205~{\AA} line blends with a \ion{V}{1}
line, while the 3819~{\AA} one is affected by a wing component of a 
strong Fe line at 3820~{\AA}. Since these Eu lines are quite weak in
the spectrum of HD~122563, in contrast to those in r-process enhanced
stars, the effect of the blending is significant.  Hence,
we adopt the result from the 4129~{\AA}. We note that the Eu
abundances derived from the excluded lines, within their relatively
large errors, agree with the value from the 4129~{\AA} line.

The effect of isotope shift for Gd lines was included in the analysis
by \citet{johnson04} for the s-process enhanced star
CS~31062-050.  This effect is, however, neglected in the present
analysis, because the Gd lines of HD~122563 are very weak.  We have
detected three Gd lines (Table \ref{tab:table1-4}). 
The Gd abundance derived from the
3481 {\AA} line is significantly higher than those from the other
two, and an upper limit estimated from another line
(3331~{\AA}). We regard this as a result of contamination of the 3481~{\AA} line by some
unidentified lines, and exclude this line in the
determination of the final Gd abundance.

Yb is the heaviest element detected in our spectrum.  We measured two
Yb lines at 3220{\AA} and 3694{\AA}.  A spectrum synthesis technique was
applied to the analysis including the effect of hyperfine splitting
for these lines (Sneden et al. 2003, private communication). 
Agreement of the results from the two lines is fairly good.

\begin{figure}[p]
\includegraphics[width=13cm]{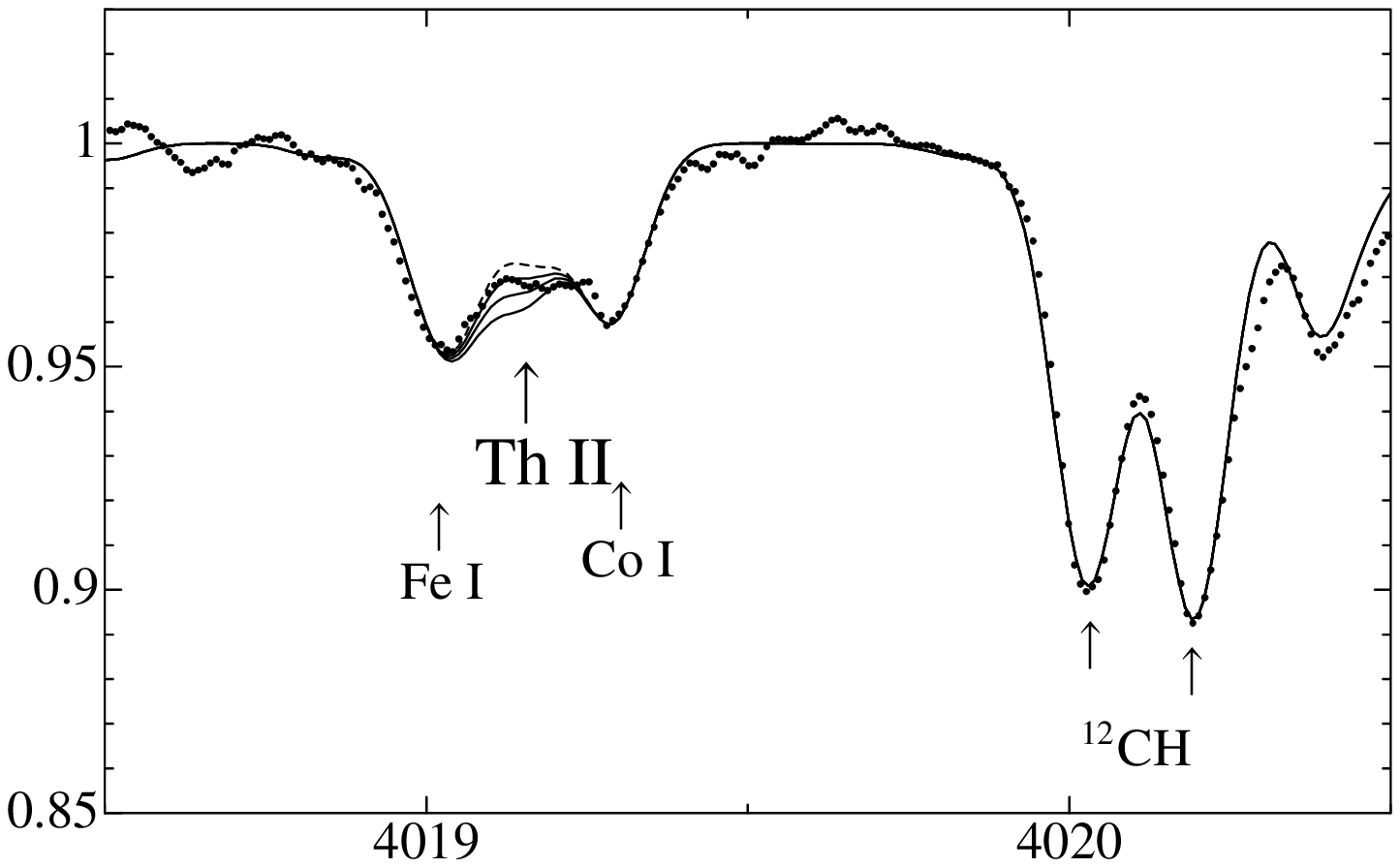}

\includegraphics[width=13cm]{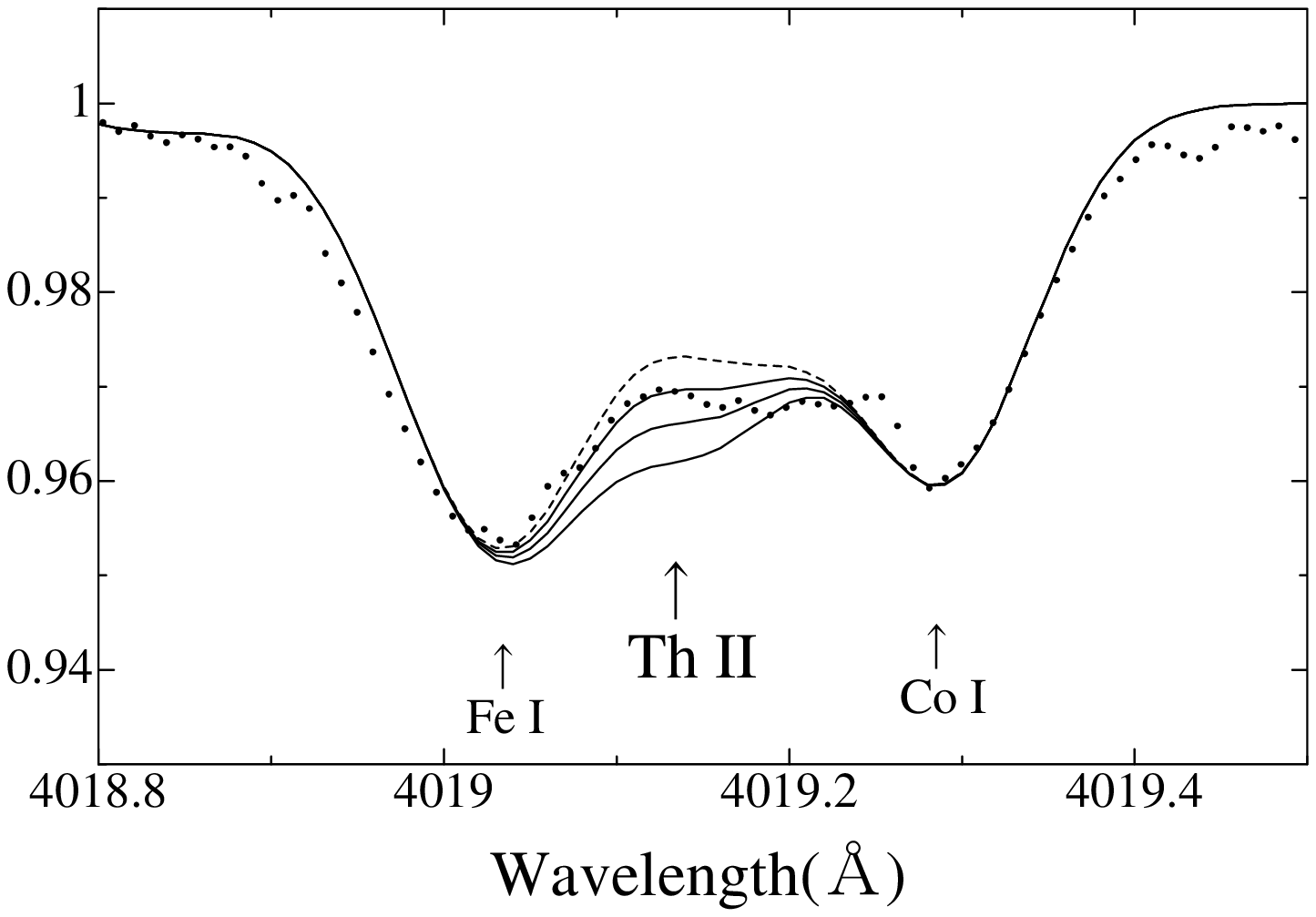}
\caption{The observed \ion{Th}{2} 4019 {\AA} line in HD~122563, and
synthetic spectra at
log$\varepsilon$(Th) $= -\infty$ (dashed line),$ -3.35, -3.05, -2.85$}
\label{fig:th}
\end{figure}

The upper limits of Ir and Th abundances are obtained from the analyses 
of \ion{Ir}{1} 3800~{\AA} and \ion{Th}{2} 4019~{\AA} line.
The line of \ion{Th}{2} 4019~{\AA} is affected by contamination by 
other lines (Figure \ref{fig:th}).
We analyzed this line using the line list of \citet{johnson01} for 
this wavelength region.
We adopted $^{12}$C/$^{13}$C = 5 to estimated the blending of $^{13}$CH lines. 
Since the blends of Fe and Co lines are severe, no clear Th feature 
is identified, although the quality of the spectrum is very high 
(S/N = 950 at 4020~{\AA}).
We derived the upper limit of the Th abundance (log $\epsilon$ (Th) $< -3.05$) 
from the fitting of synthetic spectra by eye.

\section{Comparisons with previous studies}

Neutron-capture elements have been studied for this object by many authors.  
Recently, \citet{westin00}, \citet{johnson01}, \citet{honda04}, 
and \citet{aoki05} performed detailed abundance analyses based on 
high resolution spectroscopy ($R\gtrsim 50,000$) for this object.  
\citet{burris00} also determined the abundances of eight neutron-capture 
elements, though their resolving power ($R=20,000$) is not as high as 
those in the above studies.  
Recently, \citet{cowan05} derived the abundances 
of Ge, Zr, La and Eu from the UV region (2410 $\sim$ 3070~{\AA}) 
of this object using Hubble Space Telescope, and also UV-Blue region 
(3150 $\sim$~4600~{\AA}) with Keck I HIRES.

\begin{figure}[p]
\includegraphics[width=8cm]{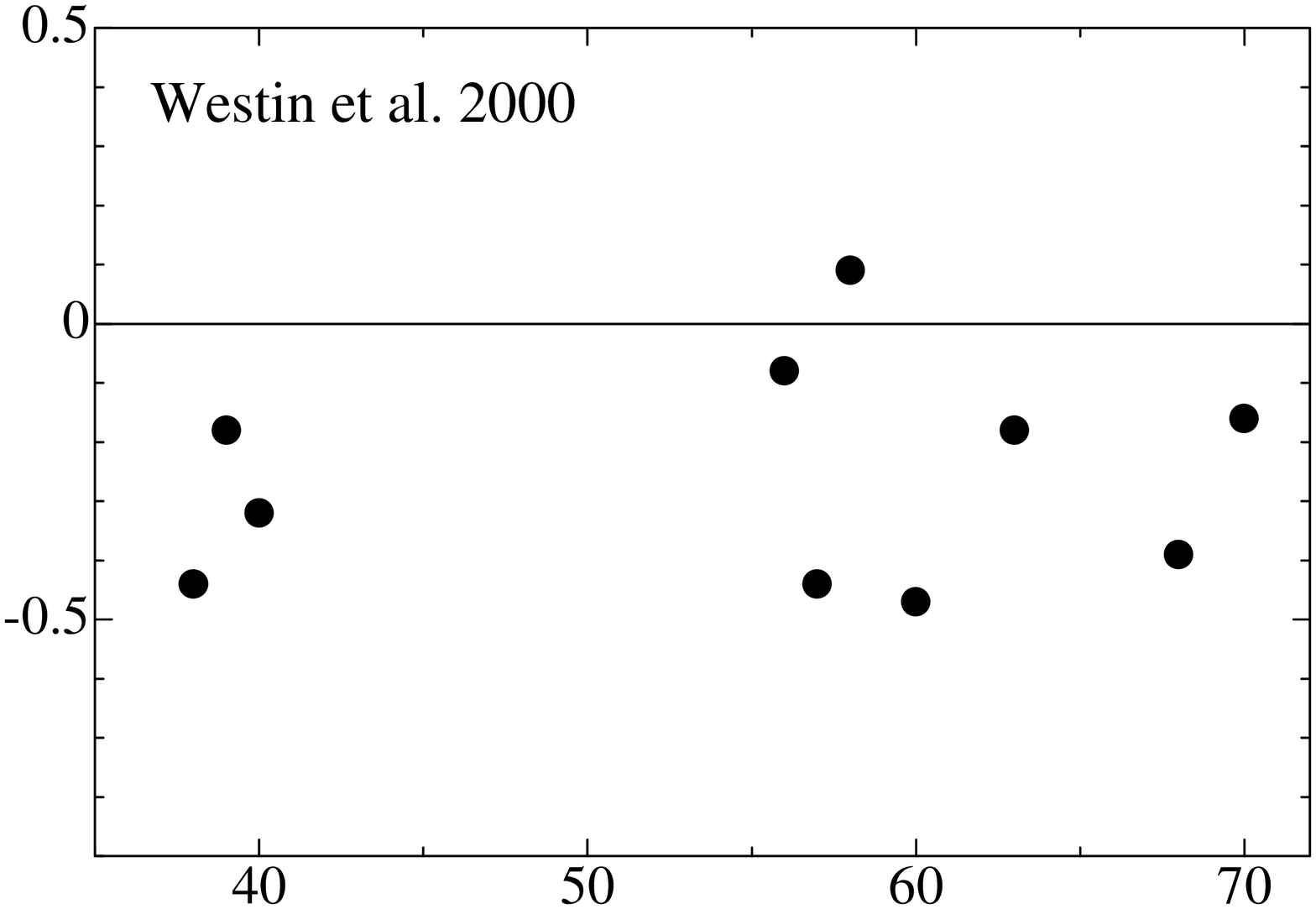}
\includegraphics[width=8cm]{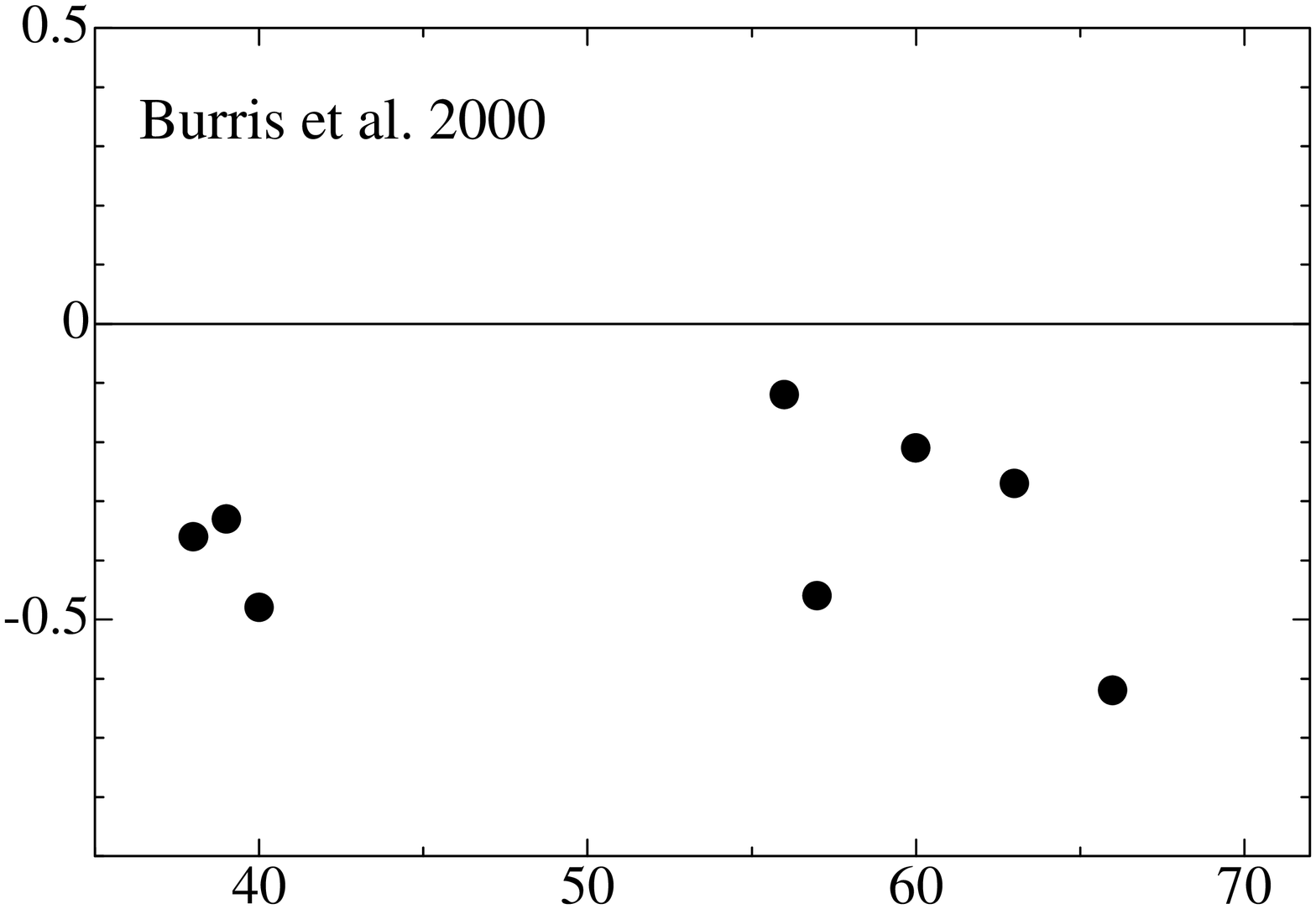}

\includegraphics[width=8cm]{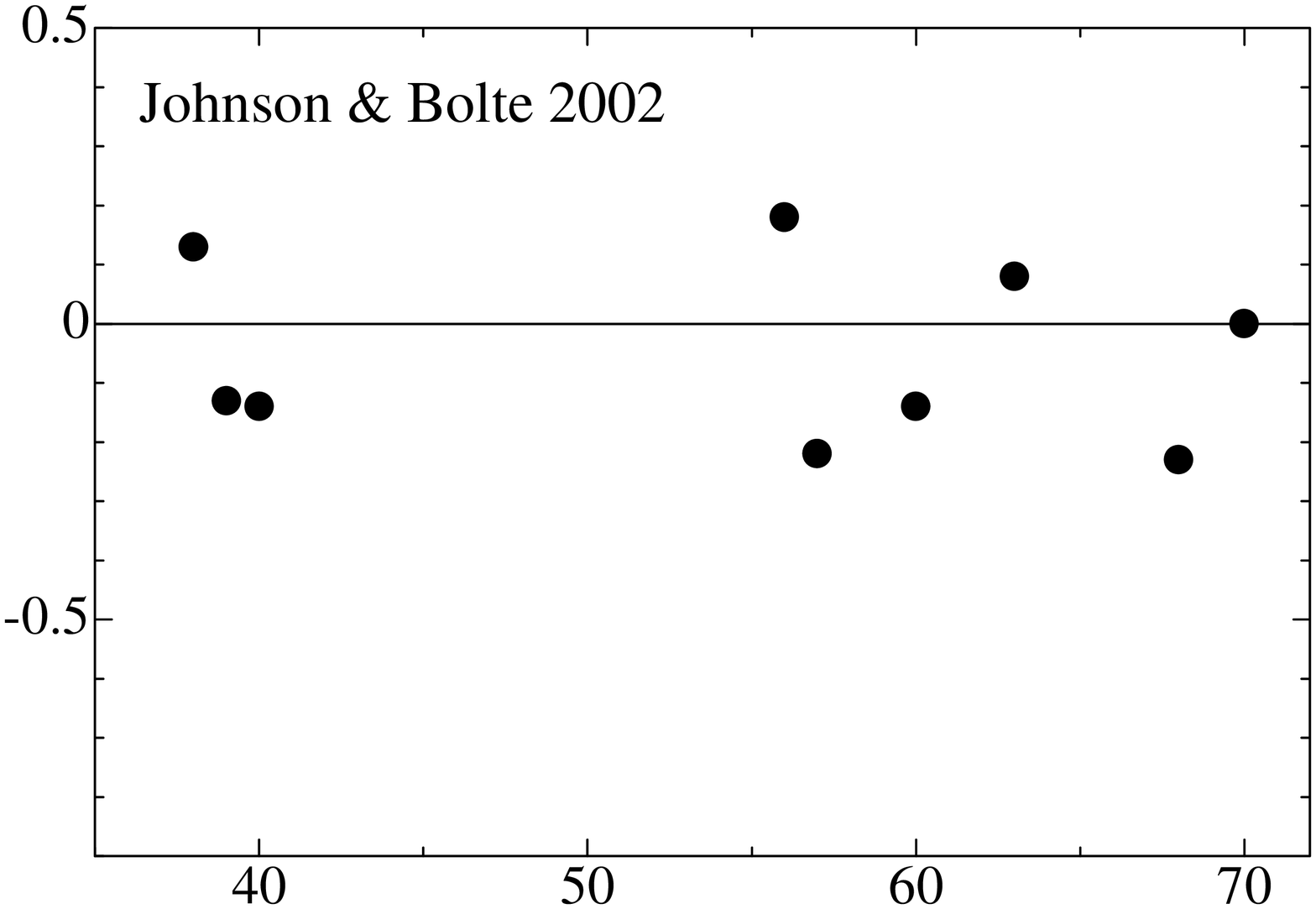}
\includegraphics[width=8cm]{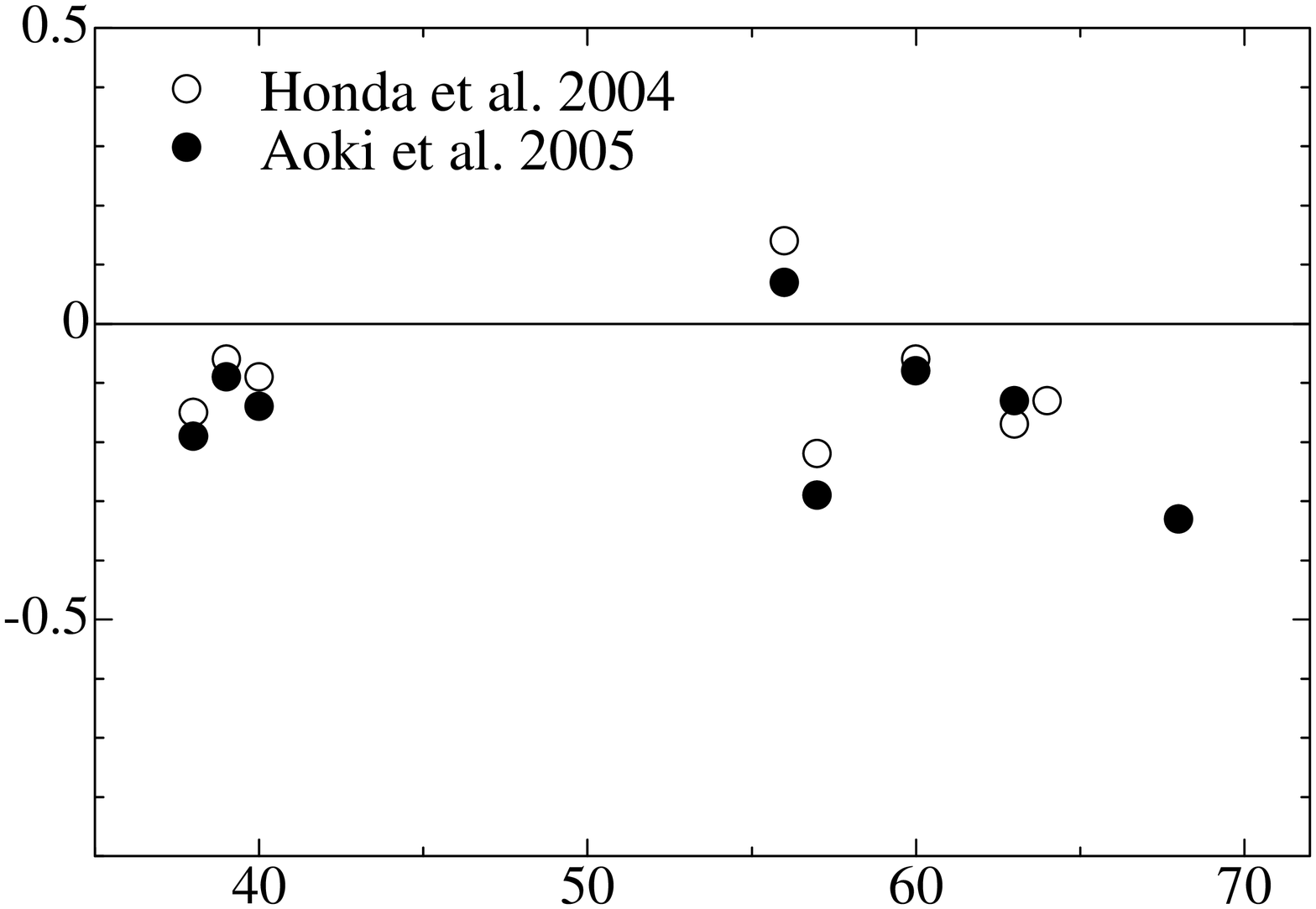}

\caption{Comparisons with the abundances derived by the present analysis and 
previous studies (Westin et al. 2000, Burris et al. 2000, Johnson \& Bolte 2001,
Honda et al. 2004 and Aoki et al. 2005) in the sense
log$\varepsilon_{\rm this}$ $-$ log$\varepsilon_{\rm other}$, 
as a function of atomic number. }
\label{fig:comp}
\end{figure}

Figure \ref{fig:comp} shows comparisons of the abundances of neutron-capture
elements derived by the present analysis with those of previous work.
Our results are in good agreement with \citet{johnson02}, \citet{honda04} 
and \citet{aoki05}.  However, the abundances of
neutron-capture elements determined by the present analysis are
systematically lower than those by \citet{westin00} and \citet{burris00}. 
Here we inspect the discrepancy between the results of
\citet{westin00} and ours in some detail, because their work
determined abundances of a larger number of elements than \citet{burris00}, 
and they reported the equivalent widths used
in the abundance analysis for some lines (unfortunately, the
equivalent widths were not given for the lines for which spectrum
synthesis technique was applied). 

The atmospheric parameters adopted in our analysis are in good
agreement with theirs: the differences (our results minus those of \citet{westin00}) 
of $T_{\rm eff}$, $\log g$, [Fe/H] and $v_{\rm turb}$ are +70~K, 
--0.2~dex, --0.07~dex and --0.3~km~s$^{-1}$, respectively.
These differences results in differences of abundances smaller than 0.1 
dex for heavy neutron-capture elements (see Table \ref{tab:error}).
Therefore, the difference of the
adopted atmospheric parameters is not a reason for the abundance
discrepancy.

We also found good agreements in the equivalent widths of
the lines reported by both works. However, the $gf$ value of the Nd
4061~{\AA} line used by \citet{westin00} is lower by 0.25~dex than
ours. Though the $gf$ values of La adopted by the two works are quite
similar, no information on the treatment of the hyperfine splitting
was given by \citet{westin00}, while the effect is fully included
in our analysis. 
We adopted recent line data for \ion{La}{2} \citep{lawler01a} 
and \ion{Nd}{2} \citep{denhartog03}, which should be more accurate than those
used by \citet{westin00}. Thus, the discrepancy between the
results of the two work might be partially explained by the difference
of the adopted line data. However, there seems to exist small
($\sim$0.2~dex) systematic differences, for which no clear reason is
identified.

\section{Discussion}

We investigated a high quality UV-blue spectrum of HD~122563, and have
detected 19 neutron-capture elements including Nb, Mo, Ru, Pd, Ag, Pr,
and Sm, which are detected for the first time by our study for this object, and
upper-limits for 5 elements including Th.  The derived abundances are
given in Table \ref{tab:table2}. 
In this section, we discuss the abundance pattern of
the neutron-capture elements, along with Cu and Zn, to investigate the
origin of neutron-capture elements in this object.

\subsection{Overall abundance pattern of neutron-capture elements}

The abundance ratio between Ba (or La) and Eu is used as an indicator
of the origin of neutron-capture elements.  The value of [Ba/Eu] of
the solar system r-process component is --0.81, while that of the
s-process component is +1.45 \citep{burris00}.  The value of
the [Ba/Eu] is --0.50 in HD~122563.  In addition, the value of [La/Eu]
of the solar system r-process component is --0.59, while that in 
HD~122563 is --0.50.  These results indicate that the heavy ($Z$ $>$ 56) 
neutron-capture elements of this object are principally associated with
the r-process, and the contribution of the (main) s-process is small if
any. We recall that this object is very metal-poor ([Fe/H]$=-2.7$)
with no excess of carbon ([C/Fe]=$-0.4$).

\begin{figure}[p]
\includegraphics[width=13cm]{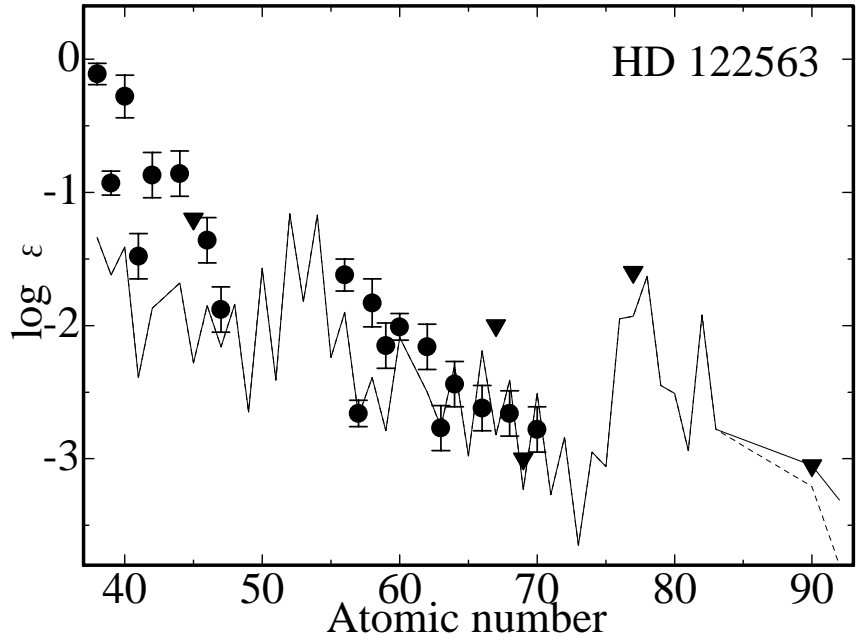}
\caption{The abundances of HD~122563 compared to the scaled solar-system r-process pattern (normalised at Eu).}
\label{fig:r}
\end{figure}

Figure \ref{fig:r} shows the abundance pattern of HD~122563, comparing 
with that of the solar-system r-process component.  
The solar system r-process abundance pattern is scaled to the Eu 
abundance of HD~122563.  
This figure clearly shows that the abundances of light neutron-capture 
elements (38 $\leq$ Z $\leq$ 47) are much higher than those of heavy 
ones in HD~122563, compared to the solar-system r-process pattern. 
This behavior is very different from that found in
r-process enhanced stars \citep[e.g. CS~22892-052, CS~31082-001:][]{sneden03,hill02}.
Although we indicated above that the heavy neutron-capture elements are
principally associated with the r-process, Figure \ref{fig:r} indicates that 
even in the range 56 $\le$ $Z$ $\le$ 59 the abundances in HD~122563 exceed 
those of the scaled solar r-process, with only La falling on the 
Eu-normalised curve. This suggests that the tendency for the light 
neutron-capture
elements (38 $\leq$ $Z$ $\leq$ 47) to fall above the scaled solar r-process 
curve is part of a general, atomic-number dependent, trend.

\begin{figure}[p]
\includegraphics[width=13cm]{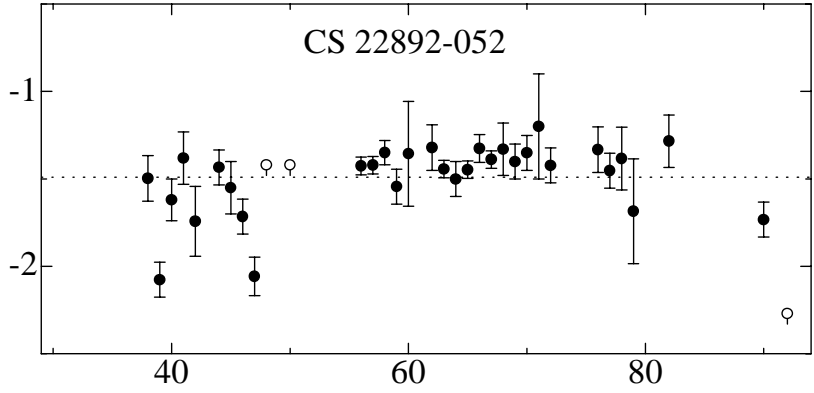}
\includegraphics[width=13cm]{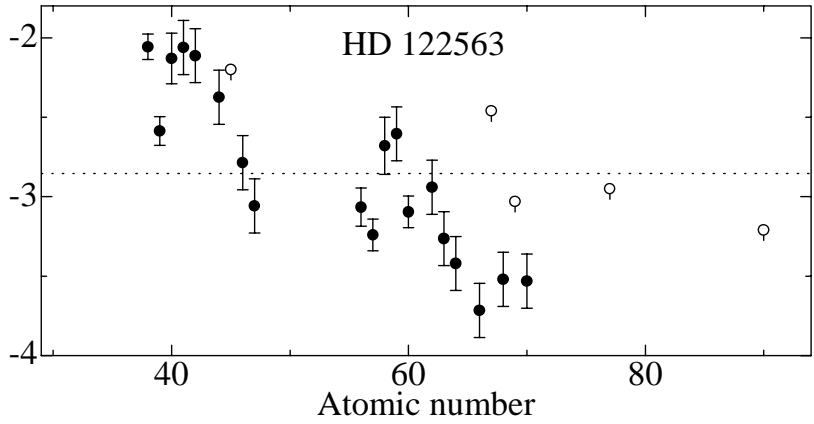}
\caption{Logarithmic differences from the solar system r-process 
pattern. Dashed line indicates the average value of difference.}
\label{fig:f7}
\end{figure}

In order to demonstrate this behavior more clearly, Figure \ref{fig:f7} 
shows the logarithmic difference of the abundances of this object from 
the solar system r-process pattern as a function of atomic number.  
We also show that of the r-process enhanced star CS~22892-052 
\citep{sneden03} in the same way.  
We find good agreement between the abundance pattern of CS~22892-052 
and the solar-system r-process one, at least for elements with 
56$\leq Z \leq76$. 
By way of contrast, our new measurements for HD~122563 clearly 
demonstrate a quite different abundance pattern from that of the 
solar-system r-process component and of CS~22892--052.

The excess of light neutron-capture elements (e.g. Sr) with respect to
the heavy ones (e.g. Ba) was known for HD~122563 by previous studies
(see section 1). However, the abundances of elements having
intermediate mass (e.g. Mo, Pd) were measured in detail for the
first time by the present work. 
The abundances of these elements are intermediate, and,
hence, the abundances of neutron-capture elements gradually and
continuously decrease with increasing atomic number from $Z=38$ to
$Z=56$. This trend is a key to investigating the nucleosynthesis process
that is responsible for neutron-capture elements in this object.

\begin{figure}[p]
\includegraphics[width=13cm]{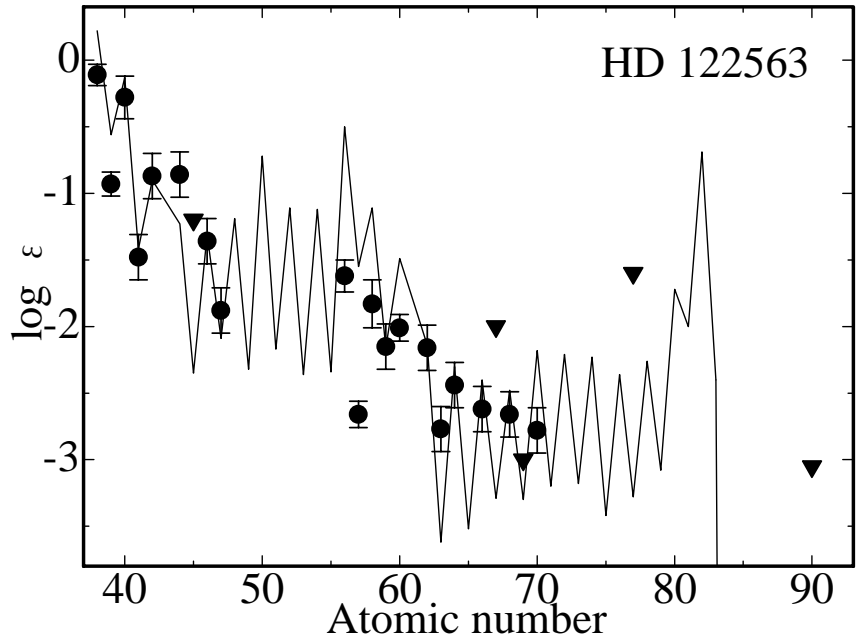}
\caption{The abundances of HD~122563 compared to the scaled solar-system s-process pattern.}
\label{fig:s}
\end{figure}

For comparison purposes, we also attempted to compare the abundances of HD~122563 with the 
solar system s-process distribution (Figure \ref{fig:s}).
In this case, Ba, La, Ce, and Eu show large deviations from the solar 
system s-process curve, although the overall abundance pattern from Sr 
to Yb seems to be similar.
Since the s-process abundance pattern in the solar-system is a result of a 
combination of some individual processes (at least the weak and main 
components of the s-process), we presume this apparent ``agreement'' is physically not important.

We also have detected four neutron-capture elements heavier than Eu
(Gd, Dy, Er, Yb: $Z \geq$ 64). The abundances of these elements also show
a decreasing trend as a function of atomic number, with respect to the
solar-system r-process pattern. This also has an impact on the
understanding of the origin of neutron-capture elements in this object. 
Th is a radioactive element and is synthesized only by the r-process. 
The upper limit of the Th abundance determined by the present work indicated 
that this object does not show an excess of Th with respect to Eu (and other 
heavy neutron-capture elements), compared to most r-process enhanced stars.

The Cu and Zn abundances of HD~122563 are typical values found 
in very metal-poor stars \citep[e.g.,][]{johnson02a,cayrel04}. 
That is, no clear difference is found in the abundances of these 
elements between the stars having high and low abundances of light 
neutron-capture elements. 
This result may give a hint for the origins of Cu and Zn, as well as 
of light neutron-capture elements, which are still unclear.

\subsection{What is the source of neutron-capture elements in HD~122563 ?}

Here we consider the origin of neutron-capture elements in this
object based on the abundance pattern determined by the present work. 

\subsubsection{Light neutron-capture elements}

The abundance ratios between light and heavy neutron-capture elements
(e.g., Sr/Ba) in HD~122563 are clearly different from those in r-process 
enhanced stars, as already mentioned in \S 1. 
Therefore, the r-process responsible
for the heavy neutron-capture elements in the Solar System (the so-called
``main'' r-process) is not an important source of, at least, light 
neutron-capture elements in HD~122563. 
The main s-process is also excluded from the
possible source of light neutron-capture elements, because that yields
even lower abundance ratios between light and heavy neutron-capture
elements at low metallicity \citep[e.g.][]{busso99,aoki02}. 

The weak component of the s-process was introduced to interpret the light
s-process nuclei in the Solar System. However, the gradually
decreasing trend of elemental abundances from Sr to Ba found in
HD~122563 does not resemble that of the weak s-process, which
predicts a rapid drop of abundances at $A\sim 90$ (i.e. Y or Zr). It
should be noted in addition that the weak s-process is expected to be inefficient 
at low metallicity, because of the lack of the neutron source $^{22}$Ne
\citep[e.g.][]{prantzos90,raiteri93} in addition to the lack of iron seeds.

Thus, another process that has efficiently yielded light
neutron-capture elements at low metallicity is required to explain the
abundance pattern of neutron-capture elements in HD~122563. The
presence of such a process has been indicated by recent observational
studies of metal-deficient stars with no excess of heavy
neutron-capture elements \citep[][and references therein]{aoki05}. 
This process was called the ``Light Element Primary Process'' (LEPP) 
by \citet{travaglio04}, who estimated its contribution to the
solar abundances. 
\citet{wanajo01} have shown that such light r-process nuclei (up to $A
\sim 130$) are produced in neutrino winds 
as a result of ``weak" (or
failed) r-processing \cite[see also][]{Wanajo05}. 
It is likely, therefore, that these light neutron-capture elements originate from
the core-collapse supernovae, 
although a contribution from other sources
cannot be excluded.
The abundance pattern of light to
heavy neutron-capture elements measured for HD~122563 provides a
unique constraint on such model calculations. 

\subsubsection{Heavy neutron-capture elements}

\begin{figure}[p]
\includegraphics[width=13cm]{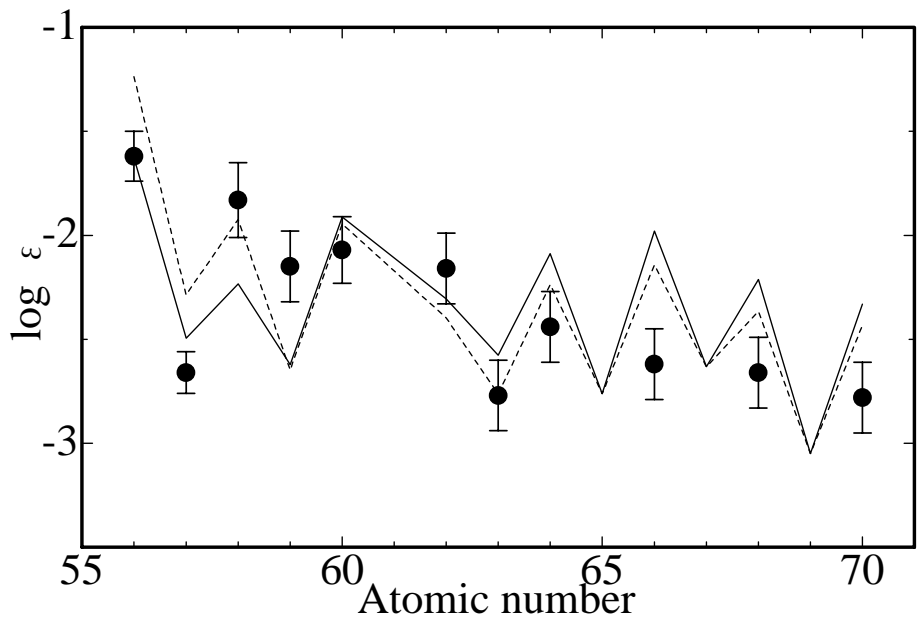}
\caption{The comparison with the scaled solar-system r-process pattern (solid line) and the best fitting r+s pattern (dashed line).}
\label{fig:r+s}
\end{figure}

The abundance pattern of heavy neutron-capture elements of HD~122563
shows a significant departure from the r-process component in
solar-system material, though the departure is not as large as that
found for light neutron-capture elements (Figure \ref{fig:r}). 
Figure \ref{fig:r+s} shows a comparison between the abundances of elements with
$56\leq Z \leq 70$ measured for HD~122563 (filled circles) and the
Solar-System r-process abundance pattern \citep{burris00}, 
which is scaled to achieve the best fit to the observed data (the solid
line). (Recall that the scaling in Figure~\ref{fig:r} was to Eu.) 
Significant disagreements are found for Ce and Pr, as well as for Dy, Er and Yb.
Given the fact that the abundance patterns of heavy elements measured 
for r-process enhanced stars show excellent agreement with that of 
the Solar-System, the different abundance pattern of HD~122563 must be significant.
Some process that produces a different abundance pattern of heavy neutron-capture 
elements from that of the main r-process is at least required to explain the abundance pattern of this star.

One possibility is to assume a small contribution of the (main) s-process.  
In order to examine this possibility, we adopted the r-process 
and s-process abundance patterns in the Solar-System
as determined by \citet{burris00}, and searched for the
combination that gives the best fit to the abundance pattern of HD~122563.
The result is shown by the dashed line in Figure \ref{fig:r+s}. 
In this case, the contributions of the s-process to Ba, La,
and Ce are significant ($\sim$ 70\%), while those to elements heavier
than Nd are small ($<$ 10\%). 
It should be noted that such a small contribution of the
 main s-process does not affect the abundance pattern of light 
neutron-capture elements discussed in the previous subsection. 

The root-mean-square (rms) of the logarithmic abundance difference 
between the observed and calculated abundance patterns is 0.29~dex for the
case that a contribution from the s-process is introduced,
compared with 0.34~dex if only the r-process is assumed. Thus, the fit
is slightly better if an s-process contribution is introduced. 
However, the improved fit comes at the cost of an additional free parameter 
(the relative contribution of the s- and r-processes), and it is not clear that
assuming an s-process contribution is justifiable from the point of view of 
the star's nucleosynthetic history.
In particular, the
discrepancy found in the abundance pattern of elements heavier than Sm
cannot be explained by this approach, because the production of such heavy
elements by the s-process is very small. 

This discrepancy leads us to consider an alternative possibility that 
the heavy neutron-capture elements of this star are also a result of 
the unidentified (single) process that is responsible for the large 
enhancement of the light neutron-capture elements discussed in \S~5.2.1. 
\citet{wanajo01} have suggested that the small (but non-negligible)
amount of heavy r-process nuclei ($A > 130$) with non solar r-process 
abundance pattern can be produced in neutrino winds if the 
entropy of the neutrino-heated matter is enough high (but smaller than that required for the main r-process), 
in addition to the light r-process nuclei.
If this is true, the abundance pattern of
heavy neutron-capture elements in this object may give another hint for,
or constraint on, the modeling of this process.

The differences between the abundance patterns of heavy neutron-capture 
elements in
HD~122563 and the solar-system r-process abundance pattern are not
large. Further observational data are clearly required to derive 
definitive conclusions. However, it would be very difficult to improve
the measurement of abundances for HD~122563, because the quality of
the spectrum used in the present work is extremely high. One
possibility is to reconsider the atomic line data to reduce the
uncertainties in abundance measurements. Another important
observational study is to apply similar analysis to other metal-poor
stars having high abundance ratios of light to heavy neutron-capture
elements. In particular, measurements for stars with lower metallicity
than HD~122563 are important, because no s-process contribution is
expected for such stars.

\section{Conclusions}

We obtained a high resolution, high S/N UV-blue spectrum of the very metal-poor 
star HD~122563 with Subaru/HDS.
The abundances were measured for 19 neutron-capture elements, among which 
seven elements are detected for the first time in this star.
Our new measurements for HD~122563 clearly demonstrate that the 
elemental abundances gradually decrease relative to the scaled solar r-process
abundances with increasing atomic number, at least from Sr to Yb.
We have considered whether the higher abundances of lighter elements can be 
explained by contributions from the weak s- or main s-processes, and find 
these not to provide satisfactory explanations.
The abundance pattern of elements with $38\leq Z \leq 47$ does not agree with any prediction of neutron-capture process models known.
The abundance pattern of a wide range of 
neutron-capture elements determined for HD~122563 
therefore provides new strong 
constraints on models of nucleosynthesis for very 
metal-poor stars, in particular those with excesses of light neutron-capture elements but without 
enhancement of heavy ones.

\acknowledgments

This work was supported in part by a Grant-in-Aid for the Japan-France
Integrated Action Program (SAKURA), awarded by the Japan Society for the
Promotion of Science, and Scientific Research (17740108) from the
Ministry of Education, Culture, Sports, Science, and Technology of
Japan.
Most of the data reduction was carried out at the Astronomical Data 
Analysis Center (ADAC) of the National Astronomical Observatory of Japan.

\begin{deluxetable}{lcccccc}
\tablewidth{0pt}
\tablecaption{LINE DATA AND EQUIVALENT WIDTHS \label{tab:table1-4}}
\startdata
\tableline
\tableline
Wavelength & L.E.P.(eV) & log$gf$ & log$\epsilon$ & {\it W}(m{\AA}) & ref \\\hline
\ion{Cu}{1}, $Z=29$ &  &  &  &  &  \\\hline
3247.53 & 0.000  & --0.060  & 1.04  & 115.1* & 8\\
3273.95 & 0.000  & --0.360  & 1.24  & 112.3  & 8\\
5105.55 & 1.390  & --1.520  & 0.66  & 3.3    & 2\\\hline
\ion{Zn}{1}, $Z=30$ &  &  &  &  &  \\\hline
3302.98 & 4.030  & --0.057  & 2.07  & 20.3* & 1 \\
3345.02 & 4.078  &   0.246  & 1.78  & 19.5* & 1 \\
4722.15 & 4.030  & --0.390  & 2.09  & 14.5  & 5 \\
4810.54 & 4.080 &  --0.170  & 2.10  & 20.5  & 5\\\hline
\ion{Sr}{1}, $Z=38$ &  &  &  &  &  \\\hline
4607.33 & 0.000  &   0.280  & --0.14  & 2.4 &  7\\\hline
\ion{Sr}{2}, $Z=38$ &  &  &  &  &  \\\hline
4077.71 & 0.000  &   0.170  & --0.18 & 163.3 & 6 \\
4215.52 & 0.000  & --0.170  & --0.03  & 155.6 & 6\\\hline
\ion{Y}{2}, $Z=39$ &  &  &  &  &  \\\hline
3327.88 & 0.410  &   0.130  & --0.98  & 49.3* & 8 \\
3549.01 & 0.130  & --0.280  & --0.97  & 49.0  & 7\\
3584.52 & 0.100  & --0.410  & --0.98  & 44.1* & 10 \\
3600.74 & 0.180  &   0.280  & --1.06  & 68.2 & 7 \\
3611.04 & 0.130  &   0.010  & --1.06  & 59.0 &  7\\
3628.70 & 0.130  & --0.710  & --0.90  & 31.8 &  10\\
3710.29 & 0.180  &   0.460  & --0.98  & 82.3 &  10\\
3747.55 & 0.100  & --0.910  & --0.98  & 22.8 & 7 \\
3774.33 & 0.130  &   0.210  & --0.91  & 79.0 &  6\\
3788.70 & 0.100  & --0.070  & --0.95  & 66.6 & 6\\
3818.34 & 0.130  & --0.980  & --0.74  & 29.4 & 6 \\
3950.36 & 0.100  & --0.490  & --0.94  & 47.4 & 6 \\
4398.01 & 0.130  & --1.000  & --0.82  & 28.0 & 6 \\
4883.69 & 1.080  &   0.070  & --0.82  & 25.0 &  6\\
5087.43 & 1.080  & --0.170  & --0.91  & 13.8 &  6\\\hline
\ion{Zr}{2}, $Z=40$ &  &  &  &  &  \\\hline
3438.23 & 0.090  &   0.420  & --0.57  & 76.8* &  7\\
3457.56 & 0.560  & --0.530  & --0.17  & 27.3 &  7\\
3479.02 & 0.530  & --0.690  & --0.34  & 16.7 &  7\\
3479.39 & 0.710  &   0.170  & --0.40  & 40.2* &  7\\
3499.58 & 0.410  & --0.810  & --0.48  & 13.6* &  7\\
3505.67 & 0.160  & --0.360  & --0.42  & 47.0* &  7\\
3536.94 & 0.360  & --1.310  & --0.33  & 7.6* &  7\\
3551.96 & 0.090  & --0.310  & --0.35  & 57.3* &  8\\
3573.08 & 0.320  & --1.040  & --0.26  & 16.4 &  7\\
3578.23 & 1.220  & --0.610  & --0.29  & 3.8* &  7\\
3630.02 & 0.360  & --1.110  & --0.28  & 12.8 &  7\\
3714.78 & 0.530  & --0.930  & --0.25  & 13.6 &  7\\
3836.77 & 0.560  & --0.060  & --0.31  & 47.3 &  6\\
3998.97 & 0.560  & --0.670  &   0.07  & 36.1 &  7\\
4050.33 & 0.710  & --1.000  & --0.09  & 11.0 &  7\\
4208.98 & 0.710  & --0.460  & --0.16  & 27.1 &  6\\
4317.32 & 0.710  & --1.380  & --0.08  & 5.3 &  6\\\hline
\ion{Nb}{2}, $Z=41$ &  &  &  &  &  \\\hline
3163.40 & 0.376  &   0.260  & --1.30  & 24.2* & 1\\
3215.59 & 0.440  & --0.190  & --1.65  & 4.5* & 6\\\hline
\ion{Mo}{1}, $Z=42$ &  &  &  &  &  \\\hline
3864.10 & 0.000   & --0.010  & --0.87  & 3.3 & 8\\\hline
\ion{Ru}{1}, $Z=44$ &  &  &  &  &  \\\hline
3498.94 & 0.000  & 0.310  & --0.90  & 4.1 &  6\\
3728.03 & 0.000  & 0.270  & --0.82  & 4.7* &  6\\\hline
\ion{Rh}{1}, $Z=45$ &  &  &  &  &  \\\hline
3692.36 & 0.000  & 0.174  & $<$-1.20  & syn &  6\\\hline
\ion{Pd}{1}, $Z=46$ &  &  &  &  &  \\\hline
3404.58 & 0.810  & 0.320  & --1.31  & 6.7* &  6\\\hline
\ion{Ag}{1}, $Z=47$ &  &  &  &  &  \\\hline
3280.68 & 0.000  & --0.050  & --1.88  & 6.5* & 6 \\\hline
\ion{Ba}{2}, $Z=56$ &  &  &  &  &  \\\hline
4554.04 & 0.000  &   0.170  & --1.76  & 95.4* & 6\\
4934.10 & 0.000  & --0.150  & --1.76  & 82.2* & 6\\
5853.70 & 0.604  & --1.010  & --1.54  & 8.5*  & 6\\
6141.70 & 0.704  & --0.070  & --1.55  & 40.6* & 6\\\hline
\ion{La}{2} $Z=57$ &  &  &  &  &  \\\hline
3794.77 & 0.240  & 0.210  & --2.35  & 5.9* & 7\\
3988.52 & 0.400  & 0.210  & --2.75  & 1.7* & 4\\
3995.75 & 0.170  & --0.060  & --2.65  & 2.1* &4 \\
4086.71 & 0.000  & --0.070  & --2.53  & 4.6* & 4\\
4123.23 & 0.320  & 0.130  & --2.70  & 2.0* & 4\\\hline
\ion{Ce}{2} $Z=58$ &  &  &  &  &  \\\hline
4222.60 & 0.120  & --0.180  & --1.90  & 2.0* & 6 \\
4523.08 & 0.520  & --0.080  & --1.62  & 1.6* & 10 \\
4539.78 & 0.330  & --0.080  & --2.03  & 1.1 &  10\\
4562.37 & 0.480  & 0.190  & --2.01  & 1.4 &  10\\
4572.28 & 0.680  & 0.290  & --1.98  & 1.1* &  8\\\hline
\ion{Pr}{2} $Z=59$ &  &  &  &  &  \\\hline
4179.40 & 0.200  & 0.480  & --2.15  & 7.8* & 7 \\
4189.48 & 0.370  & 0.380  & $<$-2.15  &  & 8 \\\hline
\ion{Nd}{2} $Z=60$ &  &  &  &  &  \\\hline
3784.25 & 0.380  & 0.150  & --2.14  & 2.0 &  9\\
3826.42 & 0.064  & --0.410  & --2.00  & 2.0* & 9 \\
4061.08 & 0.471  & 0.550  & --2.00  & 5.5 & 9 \\
4232.38 & 0.064  & --0.470  & --1.89  & 2.3* & 9 \\\hline
\ion{Sm}{2} $Z=62$ &  &  &  &  &  \\\hline
4318.94 & 0.280  & --0.270  & --2.16  & 1.8* & 7 \\
4642.23 & 0.380  & --0.520  & --2.16  & 0.8 &  7\\\hline
\ion{Eu}{2} $Z=63$ &  &  &  &  &  \\\hline
3819.67\tablenotemark{a} & 0.000  &  0.510  & --2.95  & syn & 5\\
4129.70 & 0.000  &  0.220  & --2.77  & 9.0* & 5\\
4205.05\tablenotemark{a} & 0.000  &  0.210  & --2.92  & syn & 5\\
\ion{Gd}{2} $Z=64$ &  &  &  &  &  \\\hline
3331.40\tablenotemark{a} & 0.000  & --0.140  & $<$-2.15  & syn & 7\\
3481.80\tablenotemark{a} & 0.490  & 0.230  & --1.79  & 4.0* & 7\\
3549.37 & 0.240  & 0.260  & --2.40  & 2.1 &  7\\
3768.40 & 0.080  & 0.360  & --2.48  & 3.8* &  7\\\hline
\ion{Dy}{2} $Z=66$ &  &  &  &  &  \\\hline
3460.97 & 0.000  & --0.070  & --2.42  & 3.2*  &7  \\
3531.71 & 0.000  & 0.770  & --2.81  & 8.3* & 8 \\\hline
\ion{Ho}{2} $Z=67$ &  &  &  &  &  \\\hline
3398.94 & 0.000  & 0.410  & $<$-2.00  & syn & 11 \\\hline
\ion{Er}{2} $Z=68$ &  &  &  &  &  \\\hline
3499.10 & 0.060  & 0.136  & --2.75  & 2.7 &  10\\
3692.65 & 0.050  & 0.138  & --2.57  & 4.5* &  6\\\hline
\ion{Tm}{2} $Z=69$ &  &  &  &  &  \\\hline
3701.36 & 0.000  & --0.540  & $<$--3.00  & syn & 8 \\\hline
\ion{Yb}{2} $Z=70$ &  &  &  &  &  \\\hline
3289.37 & 0.000  &  0.020  & --2.70  & 27.5* & 8\\
3694.19 & 0.000  & --0.300  & --2.86  & 13.3* & 8\\\hline
\ion{Ir}{1} $Z=77$ &  &  &  &  &  \\\hline
3220.76\tablenotemark{a} & 0.350  & --0.510  & $<$-1.12  & syn & 7\\
3800.12 & 0.000  & --1.450  & $<$-1.60  & syn & 7 \\\hline
\ion{Th}{2} $Z=90$ &  &  &  &  &  \\\hline
4019.12 & 0.000  & --0.270  & $<$--3.05  & syn &  7\\\hline
\tableline
\enddata
\tablenotetext{*}{Asterisks indicate synthesized values calculated for the abundance derived by spectrum synthesis.}
\tablenotetext{a}{Lines that were not used to derive final results.
References.-- 1.Kurucz \& Bell 1995; 2.Westin et al. 2000; 3.Lawler et al. 2001a; 4.Lawler et al. 2001b; 5.Johnson 2002; 6.Hill et al. 2002; 7.Cowan et al. 2002; 8.Sneden et al. 2003; 9.Den Hartog et al. 2003; 10.Johnson et al. 2004; 11.Lawler et al. 2004.}
\end{deluxetable}

\begin{deluxetable}{lccccc}
\tablewidth{0pt}
\tablecaption{ELEMENTAL ABUNDANCES OF HD 122563 \label{tab:table2}}
\startdata
\tableline
\tableline
species & {\it Z} & log$\epsilon$ & $\sigma$ & [X/Fe] & N \\\hline
Cu & 29 &  0.98  & 0.29  & --1.15 & 3 \\
Zn & 30 &  2.01  & 0.16  & +0.18 & 4 \\
Sr & 38 & --0.11  & 0.08  & --0.26 & 3 \\
Y  & 39 & --0.93  & 0.09  & --0.37 & 15 \\
Zr & 40 & --0.28  & 0.16  & --0.10 & 17 \\
Nb & 41 & --1.48  & 0.17  & --0.13 & 2 \\
Mo & 42 & --0.87  & 0.17  & --0.02 & 1 \\
Ru & 44 & --0.86  & 0.17  & +0.07 & 2 \\
Rh & 45 & $<$--1.20  &       & $<$+0.45 & 1 \\
Pd & 46 & --1.36  & 0.17  & --0.28 & 2 \\
Ag & 47 & --1.88  & 0.17  & --0.05 & 1 \\
Ba & 56 & --1.62  & 0.12  & --1.02 & 4 \\
La & 57 & --2.66  & 0.10  & --1.02 & 5 \\
Ce & 58 & --1.83  & 0.18  & --0.64 & 6 \\
Pr & 59 & --2.15  & 0.17  & --0.09 & 1 \\
Nd & 60 & --2.01  & 0.10  & --0.69 & 4 \\
Sm & 62 & --2.16  & 0.17  & --0.40 & 2 \\
Eu & 63 & --2.77  & 0.17  & --0.52 & 1 \\
Gd & 64 & --2.44  & 0.17  & --0.76 & 2 \\
Dy & 66 & --2.62  & 0.17  & --0.99 & 2 \\
Ho & 67 & $<$--2.00  &       & $<$+0.26 & 1 \\
Er & 68 & --2.66  & 0.17  & --0.82 & 2 \\
Tm & 69 & $<$--3.00  &       & $<$--0.23 & 1 \\
Yb & 70 & --2.78  & 0.17  & --1.09 & 2 \\
Ir & 77 & $<$--1.60  &       & $<$--0.21 & 2 \\
Th & 90 & $<$--3.05  &       & $<$--0.34 & 1 \\
\tableline
\enddata
\end{deluxetable}

\begin{deluxetable}{lccccccccccc} 
\tablewidth{0pt}
\tablecaption{ERROR ESTIMATES FOR HD~122563 \label{tab:error}}
\startdata
\tableline
\tableline
 species & \multicolumn{2}{c}{$\Delta T_{\rm eff}$} && \multicolumn{2}{c}{$\Delta \log g$} && \multicolumn{2}{c}{$\Delta$ [Fe/H]} && \multicolumn{2}{c}{$\Delta 
\xi$} \\
\cline{2-3}  \cline{5-6}    \cline{8-9}    \cline{11-12}   
 & $-100K$ & $+100K$ && $-0.3$ & $+0.3$ && $+0.5$ & $-0.5$ && $-0.5$ & $+0.5$  
\\
\tableline
\ion{Cu}{1} & $-$0.22 & $+$0.17 && $+$0.06 & $-$0.07 && $-$0.15 & $+$0.06 && $+$0.23 & $-$0.26   \\
\ion{Zn}{1} & $-$0.05 & $+$0.04 && $-$0.05 & $+$0.04 && 0.00 & $-$0.01 && $+$0.03 & $-$0.02   \\
\ion{Sr}{1} & $-$0.13 & $+$0.12 && $+$0.02 & $-$0.03 && $-$0.04 & 0.00 && $+$0.01 & 0.00  \\
\ion{Sr}{2} & $-$0.12 & $+$0.11 && $-$0.05 & $+$0.03 && $-$0.03 & $+$0.05 && $+$0.31 & $-$0.39   \\
\ion{Y}{2}  & $-$0.08 & $+$0.07 && $-$0.08 & $+$0.08 && $+$0.02 & $-$0.01 && $+$0.16 & $-$0.09  \\
\ion{Zr}{2} & $-$0.08 & $+$0.07 && $-$0.09 & $+$0.08 && $+$0.03 & $-$0.02 && $+$0.08 & $-$0.05   \\
\ion{Nb}{2} & $-$0.09 & $+$0.09 && $-$0.10 & $+$0.10 && $+$0.04 & $-$0.02 && $+$0.03 & $-$0.02   \\
\ion{Mo}{1} & $-$0.19 & $+$0.17 && $+$0.02 & $-$0.01 && $-$0.03 & $+$0.00 && $+$0.00 & 0.00   \\
\ion{Ru}{1} & $-$0.20 & $+$0.17 && $+$0.01 & $-$0.02 && $-$0.04 & $+$0.00 && $+$0.01 & 0.00   \\
\ion{Pd}{1} & $-$0.20 & $+$0.17 && $+$0.01 & $-$0.03 && $-$0.04 & $-$0.01 && $+$0.01 & $-$0.01   \\
\ion{Ag}{1} & $-$0.21 & $+$0.17 && $+$0.00 & $-$0.01 && $-$0.04 & $+$0.00 && $+$0.01 & $-$0.01   \\
\ion{Ba}{2} & $-$0.09 & $+$0.07 && $-$0.09 & $+$0.07 && $+$0.01 & $-$0.01 && $+$0.08 & $-$0.06   \\
\ion{La}{2} & $-$0.08 & $+$0.07 && $-$0.10 & $+$0.09 && $+$0.03 & $-$0.02 && $+$0.00 & $-$0.00   \\
\ion{Ce}{2} & $-$0.08 & $+$0.07 && $-$0.09 & $+$0.09 && $+$0.02 & $-$0.02 && $+$0.00 & 0.00   \\
\ion{Pr}{2} & $-$0.10 & $+$0.08 && $-$0.10 & $+$0.09 && $+$0.03 & $-$0.02 && $+$0.01 & $-$0.01   \\
\ion{Nd}{2} & $-$0.08 & $+$0.08 && $-$0.09 & $+$0.10 && $+$0.04 & $-$0.02 && $+$0.00 & 0.00   \\
\ion{Sm}{2} & $-$0.09 & $+$0.07 && $-$0.09 & $+$0.08 && $+$0.03 & $-$0.02 && $+$0.00 & $+$0.00  \\
\ion{Eu}{2} & $-$0.09 & $+$0.09 && $-$0.10 & $+$0.10 && $+$0.04 & $-$0.02 && $+$0.00 & $-$0.00   \\
\ion{Gd}{2} & $-$0.10 & $+$0.08 && $-$0.11 & $+$0.09 && $+$0.03 & $-$0.02 && $+$0.00 & 0.00   \\
\ion{Dy}{2} & $-$0.09 & $+$0.10 && $-$0.10 & $+$0.10 && $+$0.05 & $-$0.02 && $+$0.01 & $-$0.01   \\
\ion{Er}{2} & $-$0.09 & $+$0.09 && $-$0.10 & $+$0.09 && $+$0.04 & $-$0.03 && $+$0.00 & $+$0.00   \\
\ion{Yb}{2} & $-$0.09 & $+$0.08 && $-$0.10 & $+$0.10 && $+$0.04 & $-$0.02 && $+$0.03 & $-$0.02   \\
\tableline
\enddata
\end{deluxetable}

\clearpage

\clearpage

\clearpage

\clearpage

\clearpage

\clearpage

\clearpage

\end{document}